\documentclass{aa}
\usepackage{txfonts}
\usepackage{graphicx}
\usepackage{natbib}
\bibpunct{(}{)}{;}{a}{}{,}

\begin{document}

\title{Terrestrial planet formation in low eccentricity warm--Jupiter systems.}
\author{Martyn J. Fogg \& Richard P. Nelson.}
\institute{Astronomy Unit, Queen Mary, University of London, Mile
End Road, London E1 4NS.\\
\email{M.J.Fogg@qmul.ac.uk, R.P.Nelson@qmul.ac.uk}}
\date{Received/Accepted}

\abstract
{Extrasolar giant planets are found to orbit their host stars
with a broad range of semi-major axes $0.02 \le a \le 6$ AU.
Current theories suggest that giant planets orbiting at distances
between $\simeq 0.02$ -- 2 AU probably formed at larger distances
and migrated to their current locations {{\em via}} type II migration,
disturbing any inner system of forming terrestrial planets
along the way. Migration probably halts because of
fortuitously-timed gas disk dispersal.}
{The aim of this paper is to examine the effect of giant
planet migration on the formation of inner terrestrial planet
systems. We consider situations in which the giant planet
halts migration at semi-major axes in the range 0.13 -- 1.7 AU
due to gas disk dispersal, and examine the effect of including
or neglecting type I migration forces on the forming terrestrial
system.}
{We employ an N-body code that is linked to a viscous gas disk
algorithm capable of simulating: gas loss via accretion onto the
central star and photoevaporation; gap formation by
the giant planet; type II migration of the giant; optional
type I migration of protoplanets; gas drag on
planetesimals.}
{Most of the inner system planetary building blocks survive the
passage of the giant planet, either by being shepherded inward or
scattered into exterior orbits. Systems of one or more hot-Earths
are predicted to form and remain interior to
the giant planet, especially if type~II
migration has been limited, or where type~I migration has affected
protoplanetary dynamics. Habitable planets in low eccentricity
warm-Jupiter systems appear possible if the giant planet makes a
limited incursion into the outer regions of the habitable zone (HZ),
or traverses its entire width and ceases migrating at a radial
distance of less than half that of the HZ's inner edge.}
{Type~II migration does not prevent terrestrial planet formation.
There exists a wide variety of planetary system architectures that
can potentially host habitable planets.}

\titlerunning{Terrestrial planet formation in warm--Jupiter systems}
\authorrunning{M.J. Fogg \& R.P. Nelson}

\maketitle

\section{Introduction.}\label{intro}

Giant planets are thought to form in the cool, outer, regions of a
protoplanetary disk \citep[e.g.][]{pollack,papaloizou2,boss}, in
roughly the region where Jupiter and Saturn are found in our solar
system. However, numerous giant exoplanets have been found orbiting
solar-type stars well inside the approximate position of their
nebular snowline with semi-major axes from $\sim$ 3~AU down to just
a few stellar radii \citep{butler}. The most extreme examples of
these are the so-called 'hot-Jupiters', orbiting within 0.1~AU and
accounting for about a quarter of the known giant exoplanet
inventory. Planetary migration may provide the best explanation for
the presence of the hot-Jupiter population, in particular type II
migration, where the giant planet has grown massive enough to open a
gap in its protoplanetary disk and migrates inward in step with the
disk's viscous evolution \citep[e.g.][]{lin1,lin2,ward2,nelson1}.
Giant exoplanets at intermediate distances, where eccentricities can
be high, might be explained by mutual scattering
of giant planets \citep[e.g.][]{lin3,ford,papaloizou0b,marzari},
a combination of migration and scattering \citep{adams,moorhead1},
or migration along with eccentricity excitation from the disk
\citep{papaloizou1,goldreich1,ogilvie,moorhead2}.

In the case of migrating planets, the mechanism that terminates the
migration and strands exoplanets at their present orbital radii is
unknown. Migration-halting mechanisms that might work when the
planet ventures close to the central star include tidally-induced
recession caused by
the star's rotation or Roche lobe overflow and
mass loss to the star \citep{trilling}, or intrusion by the planet
into a central cavity or surface density transition in the disk,
decoupling it from the evolution of the gas
\citep{lin2,kuchner1,masset,papaloizou3}. Halting migration further
out, beyond the $\lesssim$ 0.1~AU hot-Jupiter region, may require
that giant planets form late in the lifetime of the gas disk and
hence only have time for a partial inward migration before stranding
at an intermediate distance when the gas is lost \citep{trilling}.
Disks around T Tauri stars are observed to last for $\sim$ 1 --
10~Myr \citep{haisch} but disperse over a much shorter $\sim 10^5$
year timescale \citep{simon,wolk}: a behavior that may result
primarily by accretion of gas onto the central star combined with
photoevaporative gas loss driven by the stellar UV output
\citep{clarke,alexander}. Models of this stranding mechanism
\citep{armitage1,armitage3}, which can roughly reproduce the
exoplanet semi-major axis statistics, have raised the possibility
that fortuitous disk dispersal might also explain the presence of
the hot-Jupiter population and imply that earlier formed giant
planets could have been consumed by the cental star.

If type II migration correctly accounts for the presence of the
hot-Jupiter population then these giant planets must have traversed
their inner systems at a time when gas was still present and before
the completion of terrestrial planet formation. This prompted
initial speculations that such systems would be likely to lack any
terrestrial planets within their inner few AU \citep{armitage2}, and
since hot-Jupiters are not uncommon, they have been used to infer
significant constraints on the abundance of habitable planets
\citep{ward1}, and even their galactic location
\citep{lineweaver1,lineweaver2}. This view is contradicted however
by recent models that have simulated the process of a giant planet
migrating through an interior protoplanetary disk
\citep{fogg1,fogg2,raymond2,fogg3,mandell,fogg4}. These find that
solid material is not predominantly accreted by the giant planet or
the central star; instead, solid bodies captured at interior mean
motion resonances with the giant are shepherded inward an arbitrary
distance before being randomly scattered into an external orbit. The
net result after the migration is a partitioning of most of the
original disk material into two remnants: a compacted remnant
interior to the final orbit of the giant, which typically accretes
in a short timescale to form hot-Earth or hot-Neptune planets; and
an external disk of scattered bodies. The relative predominance of
these outcomes has been shown to be sensitive to the strength of
dissipative forces operating at the time of migration
\citep{fogg1,fogg3,fogg4} with scattering becoming increasingly
prevalent in late migration scenarios when less gas is present. All
these studies concur that a scattered disk of sufficient mass to
support renewed planet formation is likely to be generated under a
variety of conditions and that terrestrial planets should be
commonplace in hot-Jupiter systems, rather than rare or absent.

One simplification common to these previous models is that the physical
mechanism that actually halts giant planet migration is not
specified or modeled. Type II migration is artificially halted when
the giant planet has reached a preset final orbit and hence is not
determined by the structure or evolution of the gas disk. Since
these models stop migration close to the central star, whilst
significant gas is still present, they appear most realistic in the
context of a central gaseous cavity halting mechanism. The examples
that come closest to implicitly assuming fortuitous gas disk
dispersal as the halting mechanism are the late scenarios of
\citet{fogg3,fogg4} where gas densities have fallen to low levels
and migration is decelerating (see Fig.~3 in \citet{fogg3} and
Fig.~\ref{figure:3} of this paper). However, the final hot-Jupiter
orbits in these papers are still artificially
imposed at 0.1~AU and are not
controlled self-consistently by the evolution of the gas.

Our previous model adopted a 1-D, viscously evolving, gas disk
algorithm which simulates accretion onto the central star, annular
gap formation in the vicinity of a giant planet, and self-consistent
type II migration; for the nebular parameters chosen, the mass of
our gas disk exponentially declined with an e-folding time of
582\,000 years \citep{fogg3}. This sort of model runs into trouble
when simulating the late stages of gas disk dispersal as it does not
reproduce a final and abrupt $\sim 10^5$ year decline that would
accord with observations. We have corrected this deficiency here by
including a photoevaporation algorithm in our code that gradually
erodes and removes mass from our gas disk. As shown by
\citet{clarke} \& \citet{alexander}, this process has little effect
on the evolution and structure of the gas disk at early times, but
comes to dominate at later times once the rate of gas loss onto the
central star due to viscous evolution
falls below the photoevaporation rate. A rapid
dispersal of the remaining gas follows, along with the cessation of
any ongoing giant planet migration.

In this paper, we report on the results of a set of scenarios where
giant planet stranding distances are no longer prescribed but which
happen when migration runs out of steam at the time of the
disappearance of the nebular gas. We therefore specifically assume
and self-consistently model \emph{fortuitous gas disk dispersal} as
the mechanism that finalizes giant planets in their post-migration
orbits. Terrestrial planetary formation in this context is of
interest because, for a hot-Jupiter to strand at $\sim$~0.1~AU, it
must form and migrate late in the lifetime of the gas disk, when gas
densities are lower and accretion in the inner system is at a more
advanced stage than previously considered. In addition, a succession
of later scenarios than this results in a succession of shorter
migrations and larger stranding distances. This has allowed us to
extend the scope of our study to model terrestrial planet growth in
those 'warm-Jupiter' systems that may have originated as the result
of a late, partial, inward migration. In this paper we define
a `warm-Jupiter' to be one orbiting with semi-major axis in
the range $0.1 < a < 2.7$ AU, where the outer limit coincides
with the snowline.

The plan of the paper is as follows. In Section 2 we outline the
additions to our model and the initial conditions of the
simulations; in Section 3 the results are presented and discussed;
in Section 4 we consider some caveats, and in Section 5 we offer our
conclusions.

\section{Description of the model.}\label{description}

We model our systems using an enhanced version of the \emph{Mercury
6} hybrid-symplectic integrator \citep{chambers1}, run as an $N +
N'$ body simulation, where there are $N$ protoplanets embedded in a
swarm of $N'$ ``super-planetesimals" -- tracer particles with masses
a tenth of the initial masses of protoplanets that act as an
idealized ensemble of a much larger number of real planetesimals and
are capable of exerting dynamical friction on larger bodies
\citep[e.g.][]{thommes1}. The central star, giant planet, and
protoplanets interact gravitationally and can accrete and merge
inelastically with all other bodies. Super-planetesimals however are
non-self-interacting but subject to a drag force from their motion
relative to the nebular gas that is equivalent to the gas drag that
would be experienced by a single 10~km radius planetesimal. Details
of these aspects of our model are given in \citet{fogg1}.

We calculate the evolution of the nebular gas using a 1-D viscous
disk model that solves numerically a modified viscous gas disk
diffusion equation that includes the tidal torques exerted by an
embedded giant planet \citep{lin1,takeuchi} and have described its
implementation in \citet{fogg3}. The gas responds by depleting over
time via viscous accretion onto the central star; opening up an
annular gap centered on the giant planet's orbit; and forming a
partial inner cavity due to dissipation of propagating spiral waves
excited by the giant planet. The back reaction of these effects on
the giant planet is resolved as torques which self-consistently
drive type II migration. We model the possible effects of type I
migration \citep{ward2,papaloizou0,tanaka1,tanaka2,cresswell1},
where a tidal interaction with the gas disk is thought to exert an
inward radial drift and strong eccentricity and inclination damping
on protoplanets of $\sim0.1-100~\mathrm{M}_\oplus$, using a simple
algorithm described in \citet{fogg4}.

As well as modeling the dynamics of gaseous volatiles in our model,
we also track the movement of presumed solid volatiles, such as
water ice and hydrated minerals, by labeling all particles with a
composition based on their original location in the disk and summing
the composition of protoplanets as they grow. We assume a crude
three-phase initial radial composition with rocky material
originating at $<$~2~AU, material similar to chondritic meteorites
between 2 -- 2.7~AU, and trans-snowline material at $>$~2.7~AU. We
do not assign an actual water mass fraction to these phases.

\subsection{Photoevaporation-driven disk dispersal.}\label{photo}

A gas disk that viscously drains onto its central star undergoes a
power law decline with a long drawn out dispersal in conflict with
observations that final dispersal occurs over a timescale that is
short compared with the disk age. \citet{clarke} showed that this
could be explained by including a model of photoevaporation of the
disk driven by the diffuse UV flux from the central star
\citep{hollenbach}. Their results show that once the accretion rate
onto the star declines to roughly equal the outer disk
photoevaporation rate, the inner disk ceases to be resupplied from
larger radii and rapidly drains onto the star. The formation of this
inner cavity then permits direct UV illumination of the outer disk
which disperses in turn in $\sim 10^5$ years \citep{alexander}.

For our purposes we need only adopt a simple parameterization of
this type of photoevaporation model and subtract from the right hand
side of our disk diffusion equation (Eq.~7 in \citet{fogg3}) an
extra term representing a disk wind:

\begin{equation}\label{dsigdt}
\dot{\Sigma}_\mathrm{w} = K
\left(\frac{r_\mathrm{g}}{r}\right)^{2.5} ,
~~ r_{\rm g} \le r \le r_{\rm out},
\end{equation}

\noindent where $\dot{\Sigma}_\mathrm{w}$ is the rate of change of
gas surface density due to photoevaporation, $r$ is radial distance,
$r_{\mathrm{out}}$ is the disk radius, and $r_\mathrm{g}$ is the
gravitational radius: the distance beyond which ionized gas can
become unbound from the star. The constant of proportionality $K$
depends directly on the disk's total photoevaporative mass loss rate
$\dot{m}_\mathrm{w}$:

\begin{equation}\label{K}
K = \dot{m}_\mathrm{w} \left(-2 \pi r_\mathrm{g}^{2.5}
\int^{r_{\mathrm{out}}}_{r_\mathrm{g}} r^{-1.5} dr \right)^{-1} .
\end{equation}

\noindent To fit with our pre-existing viscous gas disk model, we
take $r_{\mathrm{out}} = 33~ \mathrm{AU}$, and adopt $r_\mathrm{g} =
5~ \mathrm{AU}$ and $\dot{m}_\mathrm{w} = 10^{-9} ~\mathrm{M}_\odot
~\mathrm{yr}^{-1}$ which gives $K = 1.4685 \times 10^{-12}
\mathrm{g}~\mathrm{cm}^{-2}~ \mathrm{s}^{-1}$.

\begin{figure}
 \resizebox{\hsize}{!}{\includegraphics{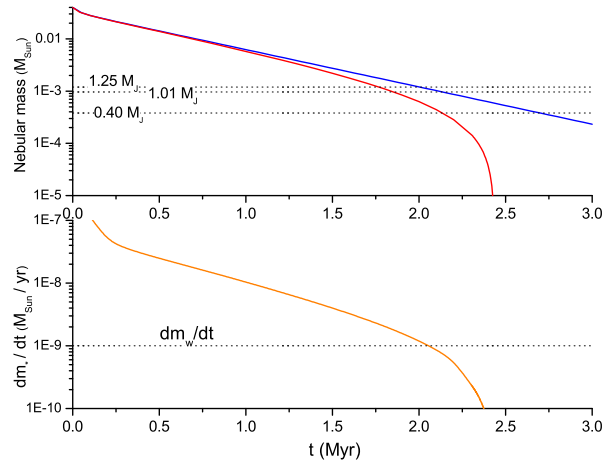}}
 \caption{\textsl{Upper panel}. Mass of the nebular gas vs. time: the blue curve
 represents our former model where mass loss occurs solely via
 viscous accretion onto the central star; the red curve
 represents our new model including photoevaporation. \textsl{Lower panel}.
 Accretion rate onto the central star $\mathrm{d}{m}_\mathrm{*} / \mathrm{d}t$.
 The dotted horizonal line represents the photoevaporation rate
 $\mathrm{d}{m}_\mathrm{w} / \mathrm{d}t = 10^{-9} ~\mathrm{M}_\odot
 ~\mathrm{yr}^{-1}$}
 \label{figure:1}
\end{figure}

In \citet{fogg3,fogg4} we assumed an initial condition of a minimum
mass solar nebula model \citep{hayashi}, scaled up in mass by a
factor of three (3$\times$MMSN), extending between 0.025 -- 33~AU
from a solar mass protostar, with an initial surface density profile
of $\Sigma_\mathrm{g} \propto r^{-1.5}$ and a total mass of
0.039~$\mathrm{M}_\odot$. Having chosen an alpha viscosity of
$\alpha = 2\times10^{-3}$, we found that, after a short lived $\sim
10^5$ year period where $\Sigma_\mathrm{g}$ close to the star
relaxes to a shallower profile, the mass of the gas disk declines
predictably with an e-folding time of 582\,000 years. This behavior
is illustrated as the upper blue curve in the top panel of
Fig.~\ref{figure:1}, which plots the nebular mass vs. time, and is
compared with the red curve which illustrates the effect of
including photoevaporation. It is evident that the two models only
diverge slowly for the first $\sim$~2~Myr, but thereafter the mass
of the photoevaporating disk drops steeply and vanishes in just a
few $\times 10^5$ years. The lower panel of Fig.~\ref{figure:1},
which plots the accretion rate onto the star
$\dot{m}_\mathrm{*}(t)$, shows that this transition in behavior
occurs around the time when $\dot{m}_\mathrm{*} \approx
\dot{m}_\mathrm{w}$.

\begin{figure}
 \resizebox{\hsize}{!}{\includegraphics{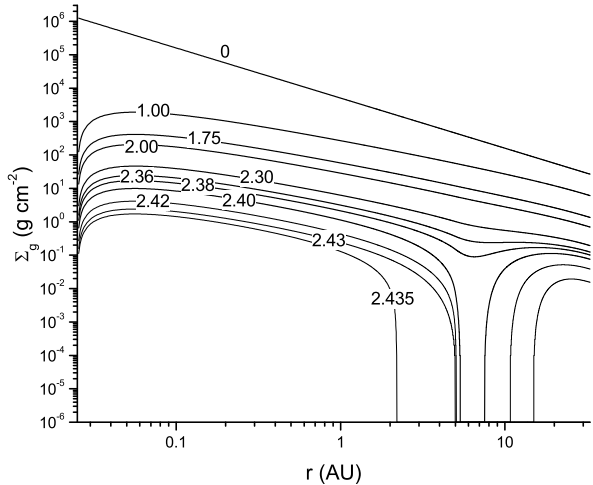}}
 \caption{Model gas disk surface density evolution. The uppermost curve
 is the initial condition; successive curves are labeled with their
 age in Myr.}
 \label{figure:2}
\end{figure}

The evolution of the gas disk surface density
$\Sigma_{\mathrm{g}}(r,t)$ is shown in Fig.~\ref{figure:2}, where
the uppermost curve represents the initial $\Sigma_\mathrm{g}
\propto r^{-1.5}$ profile and the lower curve represent successively
more evolved configurations. It can be seen that the evolution of
the nebula speeds up after $\sim$~2~Myr, with a gap at $r \approx
r_\mathrm{g}$ starting to open up at $\sim$~2.38~Myr, followed by an
accelerated decline of the inner disk thereafter. This behavior is
qualitatively similar to that described by \citet{clarke}, with the
exception that at late times our outer disk, which is truncated to a
much smaller radius, is lost even more rapidly. This makes no
difference to our mechanism for stranding giant planets as the
divergence occurs when the nebular mass has already fallen below the
level where it can drive migration. If giant planets strand well
inward of $r_\mathrm{g}$ we found that they can act against the
efficient draining of the last dregs of the inner disk onto the
central star, slightly altering the picture given in
Fig.~\ref{figure:2}. Again however, this effect is minor as it
occurs at times when the gas is very thin and does not significantly
delay the date of overall disk dispersal.

\subsection{Initial conditions and running of the simulations.}\label{initial}

We do not consider the earliest stages of planetesimal formation and
runaway growth in our modeled systems and set the $t = 0$ start date
for our simulations to be 0.5~Myr after the start of star formation.
By this time, we assume that there is no more infall of gas onto the
protoplanetary disk and the solids component of the inner disk has
reached its oligarchic growth stage, where a succession of
protoplanets, each being a few percent of an Earth-mass, have
emerged from the planetesimal swarm in near-circular orbits and with
roughly equidistant spacing in units of mutual Hill radii
\citep{kokubo}. We assume a central star of 1.0~$\mathrm{M}_{\odot}$
with initial surface density profiles for gas and solids ($\Sigma
\propto r^{-1.5}$) taken from a minimum mass solar nebular model
\citep{hayashi}, which is scaled up in mass by a factor of three
(3$\times$MMSN) to provide enough mass beyond the nebular snowline
for a giant planet to form before the loss of the nebular gas
\citep{lissauer,thommes1}. We model the gas component of the
protoplanetary disk between 0.025 -- 33~AU, giving an initial mass
of 0.039~$\mathrm{M}_\odot$, and compute its evolution with our
photoevaporating viscous gas disk algorithm, as described in the
previous section and in \citet{fogg3}. We take the alpha viscosity
of the gas to be $\alpha = 2\times10^{-3}$, which gives a viscous
evolution time at 5~AU $\simeq$ 120\,000 years.

\begin{table}
\caption{Data describing initial solids disk set-up} %
\label{table:1}  %
\centering
\begin{tabular}{c| c c| c}
 \hline\hline %
& Rocky Zone & Icy Zone& Total\\
& 0.4--2.7~AU & 2.7--5.0~AU & 0.4--5.0~AU\\
 \hline
$M_{\mathrm{solid}}$ & $9.99~\mathrm{M}_{\oplus}$ &
$24.65~\mathrm{M}_{\oplus}$ & $34.64~\mathrm{M}_{\oplus}$\\ %
 \hline
$m_{\mathrm{proto}}$ & $0.025~\mathrm{M}_{\oplus}$ &
$0.1~\mathrm{M}_{\oplus}$\\ %
$N$ & 66 & 15 & 81\\ %
 \hline
$m_{\mathrm{s-pl}}$ & $0.0025~\mathrm{M}_{\oplus}$ &
$0.01~\mathrm{M}_{\oplus}$\\ %
$N'$ & 3336 & 2315 & 5651\\
 \hline
$f_{\mathrm{proto}}$ & 0.17 & 0.06 & 0.09\\ %
 \hline\hline
\end{tabular}
\end{table}

We model the solids component of the disk initially between 0.4 -
5.0~AU and assume a snowline at 2.7~AU beyond which
the mass of solids is boosted by a
factor of 4.2 by the condensation of ices.
We generate the initial N-body components of the solids disk
in an identical manner to that detailed in \citet{fogg1} by starting
with initial protoplanetary masses of 0.025 and
0.1~$\mathrm{M}_\oplus$ interior and exterior to the snowline
respectively, spaced approximately 8 mutual Hill radii apart, with
the remainder of the material inventory consisting of
super-planetesimals with a fixed mass of 10\% of the initial masses
of the local protoplanets. Relevant data for the initial solids
components are shown in Table~\ref{table:1} which gives, for zones
interior and exterior to the snowline, values for the total mass of
solid material $M_{\mathrm{solid}}$, the number and mass of
protoplanets $N$ and $m_{\mathrm{proto}}$, and the number and mass
of super-planetesimals $N'$ and $m_{\mathrm{s-pl}}$. The parameter
$f_\mathrm{proto}$, at the foot of Table~\ref{table:1}, is the mass
fraction of the solids disk contained in protoplanets and we use
this here as a rough measure of the evolution of the disk, taking
$f_\mathrm{proto} = 0.5$ to denote the transition between oligarchic
and chaotic, or `giant impact', growth regimes \citep{goldreich2}.

\begin{figure}
 \resizebox{\hsize}{!}{\includegraphics{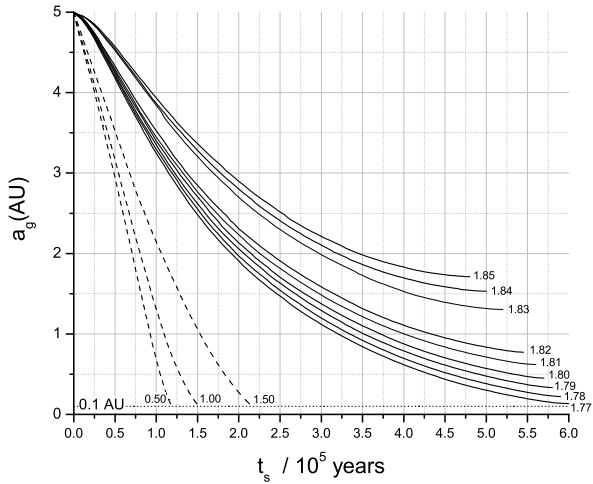}}
 \caption{Migration and stranding of $0.5~\mathrm{M_J}$ giant planets
 in our present and previous models. Giant planet semi major axis in
 AU is plotted against $t_\mathrm{s}$,
 time normalized to the start of migration, in units of $10^5$~yr.
 Black curves show the behavior of giant planets stranding naturally
 due to gas disk photoevaporation. Numeric labels are launch dates:
 the respective disk ages $t/\mathrm{Myr}$ at which the giant planet
 is introduced. Dashed curves show giant planet migrations in
 Scenarios III, IV, and V from \citet{fogg3}, where migration was
 artificially halted at 0.1~AU.}
 \label{figure:3}
\end{figure}

Our previous approach was to run our combined N-body and gas disk
model, in the absence of a giant planet, from $t = 0$ to a set of
durations distributed between 0.1 - 1.5~Myr, in order to mature the
disk to different ages and to generate a set of migration scenarios.
At the end of each of these maturation runs, we then introduced a
$0.5~\mathrm{M_J}$ giant planet at 5~AU, after removing
$0.4~\mathrm{M_J}$ of gas from a local disk annulus to provide for
the giant planet's envelope. This gas is assumed to have accreted on
top of a $0.1~\mathrm{M_J}$ solid core, composed of material
deriving from beyond the outer boundary of our modelled solids disk.
Re-starting the run at this point resulted in an inward type II
migration of the giant planet which was allowed to continue until it
reached our prescribed stranding radius of 0.1~AU. This approach is
not appropriate here as we are specifically assuming gas disk
dispersal as the stranding mechanism and therefore need to mature
the disk to between the boundaries of a temporal window within which
our giant planet can both accrete sufficient gas for its envelope
and cease migration at $\gtrsim$~0.1~AU. We located the lower limit
of this window, the age where a $0.5~\mathrm{M_J}$ giant planet
introduced at 5~AU will naturally cease migrating and come to rest
at $\sim$~0.1~AU, via experiments with the model and found it to be
$t \approx 1.77$~Myr. By this time, the mass of the nebula has
fallen to $M_{\mathrm{gas}} \approx 1.25~\mathrm{M_J}$ which is
shown as the upper dotted line in Fig.~\ref{figure:1}. The maximum
possible upper age limit of the stranding window would be when
$M_{\mathrm{gas}} = 0.4~ \mathrm{M_J}$, at $t \approx 2.13$~Myr (see
the lower dotted line in Fig.~\ref{figure:1}), giving a window
duration of $\sim 17\%$ of the simulated disk lifetime. However,
this limit would require the unrealistic condition of all the
remaining nebular gas being accreted by the giant planet. Since we
do not simulate the process of gas accretion onto the giant planet's
core, we have arbitrarily restricted the upper age limit of the
stranding window to $t = 1.85$~Myr (see the middle dotted line in
Fig.~\ref{figure:1}) by which time the mass of the nebula has fallen
to $M_{\mathrm{gas}} = 1.01~\mathrm{M_J}$, reducing the window
duration to $\sim 4\%$ of the simulated disk lifetime. We note that
the model of \citet{armitage1} predicts a stranding window of
duration $\approx 20\%$ of the disk lifetime. This longer duration
compared to ours is because they adopted a more slowly evolving disk
model and were simulating the stranding of more massive giant
planets which migrate more slowly. Clearly, variation of the many
free parameters in a model such as this can produce a variety of
stranding behaviors in simulations, but we have not considered these
here as our main focus is on the effect that the giants' stranding
distances have on the partitioning of the inner system disk and
subsequent terrestrial planet formation.

The behavior of $0.5~\mathrm{M_J}$ giant planets launched into our
model disks, between the lower and upper age limits discussed above,
is illustrated by the solid curves in Fig.~\ref{figure:3} and shows
that stranding takes place between 0.13 -- 1.71~AU. Also illustrated
by the dashed curves in Fig.~\ref{figure:3} are the migration
trajectories of the giant planets in Scenarios III, IV, and V of
\citet{fogg3} where migration takes place in a younger disk and was
artificially halted at 0.1~AU by a presumed inner disk cavity. The
fastest of these (launched at 0.5~Myr) takes place at a time when
the gas disk is still quite massive and completes its migration in
the viscous evolution time of $\sim$~120\,000 years. Later scenarios
entail longer migration times as the gas mass progressively declines
and is less effective at driving migration. In order for giant
planets to strand naturally, our present models require still later
launch times, in a context where photoevaporation is starting to
have a significant influence on the disk. Fig.~\ref{figure:3} shows
that migration speeds are considerably slower and decelerate
steadily until migration ceases after $t_s = 600\,000$ years for the
farthest travelling planet and 470\,000 years for the planet that
migrates the least. Migration in all these cases halts at $t \approx
2.4$~Myr.\footnote{Note that the gap between the $t = 1.82$~and
1.83~Myr curves is because of the substantial depletion of disk gas
at these late times, dictating the removal of the quantity required
for the giant planet's envelope from a wider annulus of the disk
than previously.}

\begin{figure*}
 \sidecaption
 \includegraphics[width=12cm]{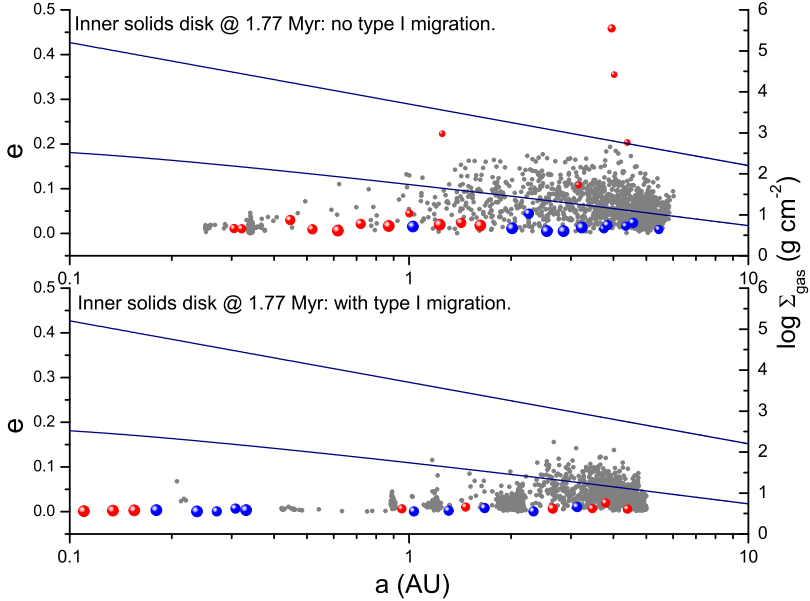}
 \caption{Eccentricity vs semi-major axis for matured solids disks at
 1.77~Myr for no type I migration (top panel) and with type I migration
 (bottom panel). Grey dots are super-planetesimals; red and blue
 circles are protoplanets originating interior and exterior to the
 snowline respectively. Gas densities are read on the right hand axes:
 the upper blue lines show gas density at $t = 0$ and the lower cyan
 lines are the current densities.}
 \label{figure:4}
\end{figure*}

Thus, we generated the scenarios for this paper by running the model
from its initial condition, without a giant planet present, to
mature the protoplanetary disk to a minimum of $t = 1.77$~Myr and
then in successive 10\,000 year increments to $t = 1.85$~Myr. Given
that the reality of strong type I migration is controversial, and to
bracket the range of possibilities, two parallel sets of scenarios
are generated: one with no type I migration forces (Run Set
\textbf{A}) and the other with type I migration and eccentricity and
inclination damping set at the maximum rate determined by
Eqs.~1~\&~2 in \citet{fogg4} (Run Set \textbf{B}). To accommodate
this total of 18 simulations, a change from our previous scenario ID
system is also required. Thus, Roman numerals are substituted with
Arabic numerals and labelling is generated by the following formula:
$\mathrm{Scenario~ID} = 1 + (t / \mathrm{Myr} - 1.77) / 0.01$. This
gives Scenarios~1,~2,~..,~9 for $t = 1.77,$~1.78,~..,~1.85~Myr. Run
set \textbf{B} which includes type~I migration is denoted by an I
subscript: e.g
Scenarios~$1_\mathrm{I},~2_\mathrm{I},~..,~9_\mathrm{I}$. During
these maturation runs the simulation inner edge was set at 0.1~AU
and any material passing interior to this boundary was eliminated
and assumed to be consumed by the central star. The configuration of
these matured solids disks at $t = 1.77$~Myr, the opening of the
stranding window, are shown in Fig.~\ref{figure:4} and
Table~\ref{table:2} gives relevant data including values for the
remaining total mass $M_{\mathrm{solid}}$, the maximum
protoplanetary mass $m_{\mathrm{max}}$, the numbers of surviving
protoplanets and super-planetesimals $N~\mathrm{\&}~N'$, and the
protoplanet to whole disk mass fraction $f_\mathrm{proto}$. Since
the scenarios of the present work start much closer together in time than
those of our previous models (0.01~Myr \emph{vs.} 0.15 -- 0.5~Myr), the
state of the solids disks generated for Scenarios~2 -- 9 does not
change greatly from that of Scenario~1.

\begin{table}
\caption{Matured solids disk at the start of the stranding window} %
\label{table:2}  %
\centering
\begin{tabular}{c| c c}
 \hline\hline %
$t$~Myr & 1.77 & 1.77 \\
Type I migration ? & no & yes \\
 \hline
$M_{\mathrm{solid}}$ & $33.52~\mathrm{M}_{\oplus}$ &
$30.42~\mathrm{M}_{\oplus}$ \\
$m_{\mathrm{max}}$ & $2.67~\mathrm{M}_{\oplus}$ &
$1.62~\mathrm{M}_{\oplus}$ \\
$N$ & 27 & 19 \\
$N'$ & 1854 & 1809 \\
$f_\mathrm{proto}$ & 0.62 & 0.52 \\
 \hline\hline
\end{tabular}
\end{table}

\begin{figure}
 \resizebox{\hsize}{!}{\includegraphics{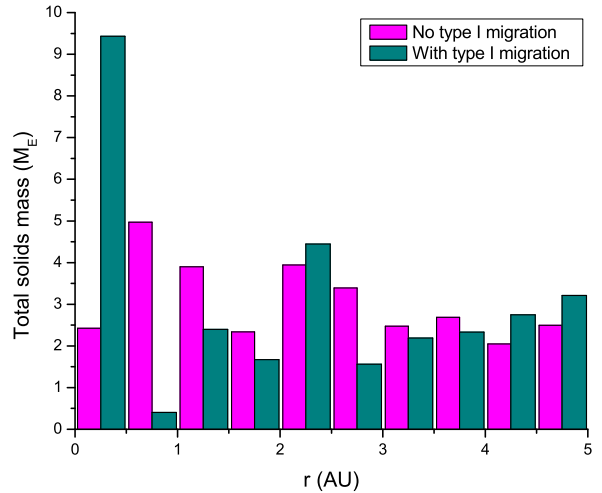}}
 \caption{Total solids mass in 0.5~AU width bins at $t = 1.77$~Myr
  for maturation runs with no type I migration (magenta bars) and with
  type I migration (cyan bars).}
 \label{figure:5}
\end{figure}

It can be seen, when comparing with the initial condition data in
Table~\ref{table:1}, that planetary growth has been strong,
especially where no type I migration is operating. In this case (the
upper panel in Fig.~\ref{figure:4}), mergers have reduced
protoplanets to a third of their former number, $m_{\mathrm{max}}$
is high (there being a $2.67~\mathrm{M}_{\oplus}$ planet present at
2.01~AU), and $f_\mathrm{proto}$ indicates that the accretion
pattern of the disk has progressed way beyond oligarchic growth into
the chaotic growth regime. Very little mass has been lost interior
to 0.1~AU ($\sim 3\%$) via dynamical spreading and gas drag induced
orbital decay of planetesimals. With type I migration, there is in
play an additional preferential damping and inward migration of the
most massive protoplanets (clearly visible in the lower panel in
Fig.~\ref{figure:4}) resulting in the loss of $\sim 12\%$ of the
disk mass beyond the simulation inner edge. This loss is mostly in
the form of large bodies as can be inferred from the lower values of
$m_{\mathrm{max}}$, $N$, and $f_\mathrm{proto}$. It might be thought
that this loss is quite modest considering our inclusion of type I
migration forces. However, in a rapidly dispersing gas disk model
such as ours, inward type I migration, which is proportional to
planetary mass, is limited at early times by the small size of
protoplanets, and at late times by low gas densities \citep[see
also][]{mcneil,daisaka}; Fig.~\ref{figure:4} shows that at $t =
1.77$~Myr, $\Sigma_\mathrm{g}$ interior to 1~AU has fallen by two
orders of magnitude. The effect of Type I migration on the radial
distribution of solids disk mass is shown in Fig.~\ref{figure:5}
where the total solids mass for both models at $t = 1.77$~Myr is
plotted in 0.5~AU width bins against radial distance. It can be seen
that beyond $\sim 2$~AU the radial mass profile of the disks in the
two models remains similar, but interior to 2~AU the more rapid pace
of protoplanetary growth has resulted, in the type I migration case,
in an inward displacement of mass caused mainly by a fractionation
of the most massive bodies from the rest of the swarm which are now
crowding the inner 0.5~AU of the system. It was shown in
\citet{fogg4} that hot-Earth type planets are more likely to accrete
and survive when a giant planet migrates through such a solids disk,
where previous type I migration has caused a radial contraction of
the inner mass distribution.

The matured disks summarized above, aged in 10\,000 year stages from
$t = 1.77$~Myr, are used as the basis for the type II giant planet
migration scenarios presented here. In each case, a giant planet of
0.5~$\mathrm{M_J}$ is inserted at 5~AU after removing
0.4~$\mathrm{M_J}$ of gas from the disk and the inner boundary of
the simulation is reset to 0.014~AU~$\cong 3~\mathrm{R}_{\odot}$,
which is approximately the radius of a solar mass T-Tauri star
\citep{bertout}. The giant planet then proceeds to clear an annular
gap in the gas and undergoes inward type II migration. The
simulations are halted when the giant planet strands as a result of
the near complete loss of the disk gas. In practise, since our
viscous disk algorithm requires a finite amount of gas in each cell
to remain stable, we assume that the giant halts inward migration
when the migration rate falls below - 0.2~$\mathrm{cm~s^{-1}} \cong
-~4.2\times10^{-7}~\mathrm{AU~yr^{-1}}$. By this time the total gas
remaining in the entire modeled disk is $< 10^{-5}~
\mathrm{M}_\odot$ and any error in stranding radius caused by this
procedure is only on the order of $\sim10^{-3}$~AU. The symplectic
time-step for these runs was set to one tenth the orbital period of
the innermost object which was achieved by dividing each simulation
into a set of sequential sub-runs with the time-step adjusted
appropriately at each restart. Since these new simulations involved
a migration of roughly triple the simulated duration of our previous
models (see Fig.~\ref{figure:3}), and since small time steps were
usually needed during the long drawn out `end game' when the giant
planet and the shepherded fraction of the solids disk are close to
their final positions, these runs took a particularly long while to
complete, requiring 4 -- 6 months of 2.8~GHz-CPU time each.

\section{Results of the model.}

\subsection{System configurations at the stranding
point.}\label{modIVres}

\begin{figure*}
\centering
  \includegraphics[width=17cm]{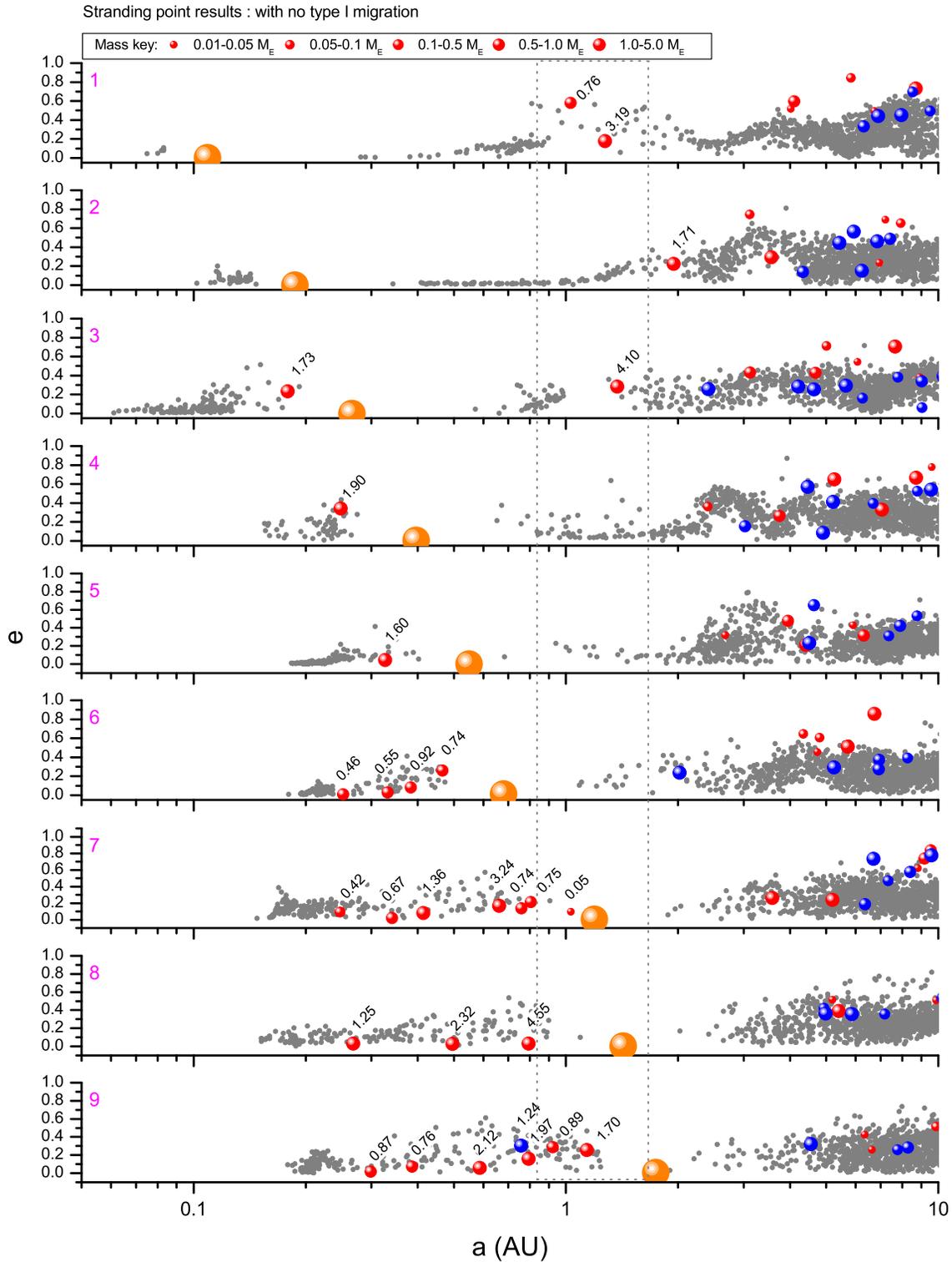}
  \caption{Run Set \textbf{A}. End points of scenarios that \emph{exclude} type~I
 migration, when the giant planet strands at its
 final semi-major axis. Eccentricity is plotted vs. semi-major
 axis with symbols colour coded as in Fig~\ref{figure:4} and
 sized according to the mass key. Scenario ID is given at the
 top left of each panel. Protoplanets interior to the giant, or
 within 1 -- 2~AU, are labelled with their mass in
 $\mathrm{M}_\oplus$. The dotted box shows the habitable zone.}
  \label{figure:6}
\end{figure*}

\begin{figure*}[!]
\centering
  \includegraphics[width=17cm]{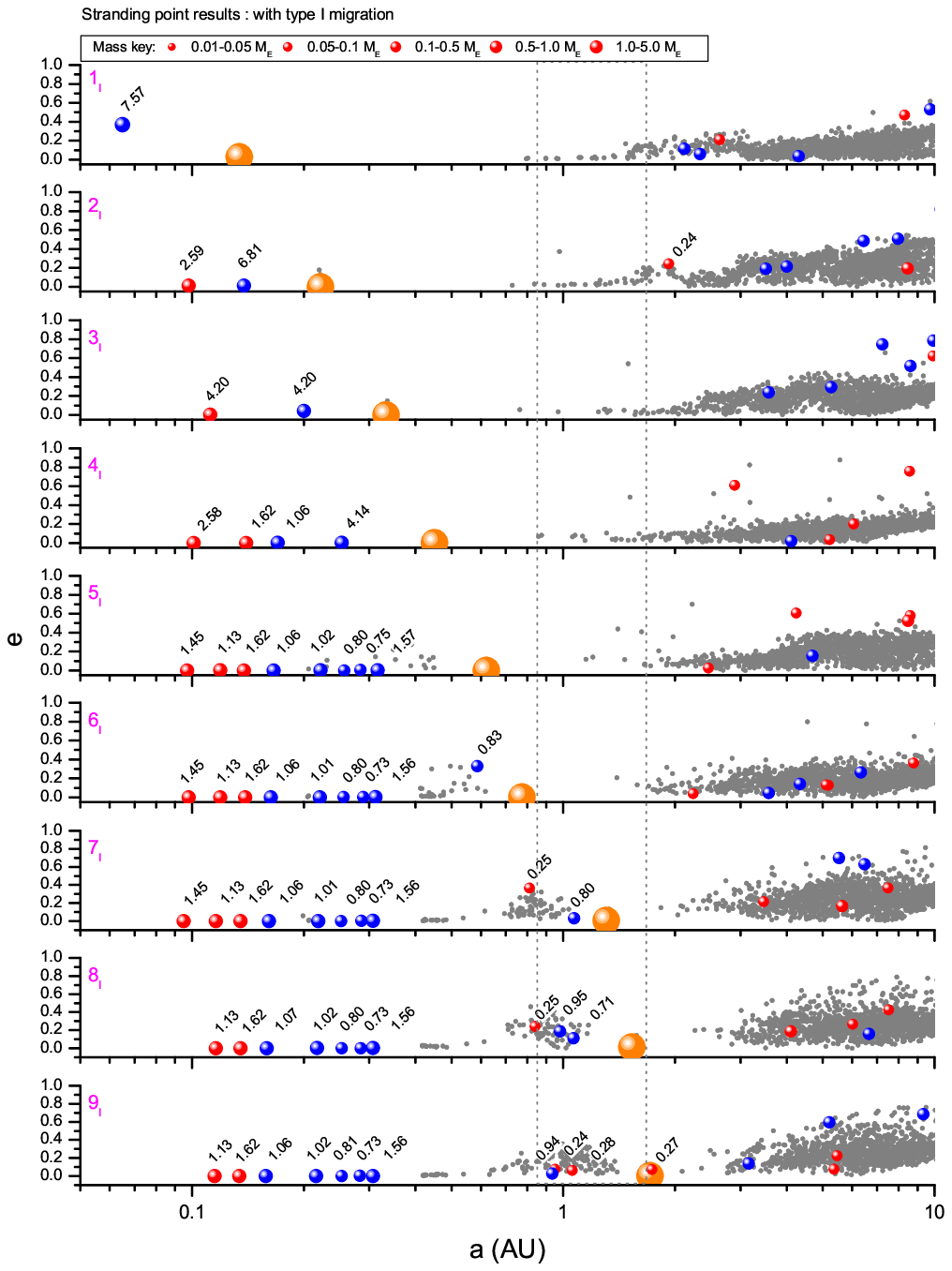}
  \caption{Run Set \textbf{B}. End points of scenarios that \emph{include} type~I
 migration, when the giant planet strands at its
 final semi-major axis. Eccentricity is plotted vs. semi-major
 axis with symbols colour coded as in Fig~\ref{figure:4} and
 sized according to the mass key. Scenario ID is given at the
 top left of each panel. Protoplanets interior to the giant, or
 within 1 -- 2~AU, are labelled with their mass in
 $\mathrm{M}_\oplus$. The dotted box shows the habitable zone.}
  \label{figure:7}
\end{figure*}

All scenarios, when run to the point at which the giant planet
ceases migrating, exhibit a varying mix of the same shepherding and
scattering effects on the solids disk shown in
\citet{fogg1,fogg3,fogg4}. It is unnecessary therefore to repeat the
previous procedure of giving a detailed account of the evolution of
one representative case. The big difference is that the scenarios
presented here result in a more restricted type~II migration, with
the giant planet stranding between semi-major axes of $a_\mathrm{g}
\approx$~0.1 -- 1.7~AU, in a context where all the damping forces
that are dependent on the disk gas are close to their minimum
possible values.

The end points\footnote{The phrase `end point' here refers to the
time at which type~II migration of the giant planet ceases due to
gas disk dispersal.} of Scenarios~1 -- 9, those \emph{without}
type~I migration (Run Set \textbf{A}), are all illustrated in
Fig.~\ref{figure:6}. Their counterparts, Scenarios~$1_\mathrm{I}$ --
$9_\mathrm{I}$, \emph{with} type~I migration operating (Run Set
\textbf{B}), are illustrated in Fig.~\ref{figure:7}. Comparison of
the figures shows the familiar partitioning of the solids disk into
shepherded interior and scattered exterior fractions, systematically
truncated by the extent of the traverse of the giant planet. Early
scenarios (1 -- 3 and $1_\mathrm{I}$ -- $3_\mathrm{I}$) are the
closest to previous models and result in relatively better populated
exterior disks and sparser interior disks without type~I migration
in force, and the opposite tendency with type~I migration. This is
in accord with the previous findings of \citet{fogg3} and
\citet{fogg4} respectively.

No surviving protoplanets are found in or close to the system's
maximum greenhouse habitable zone \citep[$\sim$~0.84 --
1.67~AU;][]{kasting} when the giant strands between $0.4 \lesssim
a_\mathrm{g} \lesssim 1.2$~AU. When stranding occurs at
$a_\mathrm{g} \lesssim 0.4$~AU, late scattered protoplanets can find
themselves emplaced in the exterior disk at distances of
$\lesssim$~2~AU and are hence candidates for evolving into future
habitable planets. However, this eventuality appears less likely if
type~I migration is influential on the dynamics, as interior disk
fractions evolve closer to the star, are better damped, and hence
are less likely to lose their contents via late scattering. When
stranding occurs at $a_\mathrm{g} \gtrsim$~1.2~AU, protoplanets are
found to survive in the inner regions of the HZ at $\sim$~1~AU in
both scenario sets. The reason that interior HZ planets are found
much closer to the giant planet than those in external orbits is
simply a reflection of the asymmetry between shepherding and
scattering behaviours. Shepherding of the interior population occurs
between the 2:1 and 4:3 resonances, causing material to accumulate
via disk compaction between $a \approx 0.63 - 0.83~a_\mathrm{g}$. In
contrast, scattering typically results in the expulsion of a
protoplanet into the exterior disk with an initial $e \gtrsim 0.5$
and periastron $\approx a_\mathrm{g}$ at the time of scattering;
hence, exterior HZ planets are usually found with semi-major axes
much larger than the final semi-major axis of the giant.
\emph{Habitable planet candidates can therefore be expected if a
migrating giant planet makes a limited excursion into the HZ.
However, if it traverses the HZ, such candidates are only expected
to be common if the giant continues its migration to a radial
distance of less than half that of the inner edge of the HZ.}
Whether the potential habitable planets visible in
Figs.~\ref{figure:6}~\&~\ref{figure:7} can survive the long final
phase of accretion that remains to be played out in their respective
systems is examined in Sect.~\ref{outerplanets}.

The interior disks that result after the giant planet's migration
stalls show a considerable difference when the two figures are
compared. When type~I migration operates (Fig.~\ref{figure:7}),
interior partitions tend to be more massive and have cleared almost
all of their planetesimal population. Hence $f_{\mathrm{proto}}
\approx 1$, but large numbers of protoplanets are also found, as
type~I eccentricity damping exerted by the residual gas acts to
reduce the effects of mutual scattering, causing protoplanets to
dynamically settle into stable resonant convoys with closely spaced,
and near-circular, orbits (originally described by \citet{mcneil}).
For example, the inner eight protoplanets at the end points of
Scenarios~$5_\mathrm{I} - 7_\mathrm{I}$, from the inside out, are
locked into a 4:3, 5:4, 4:3, 3:2, 5:4, 7:6, 7:6 configuration of
mean motion resonances. Oligarchic growth therefore ends in these
interior systems when the planetesimal field is accreted, but giant
impact growth is delayed, at least for as long as the disk gas
persists. When no type~I migration operates (Fig.~\ref{figure:6}),
interior partitions are relatively depleted of mass and contain
fewer protoplanets, in more excited orbits, alongside a surviving
population of planetesimals ($f_{\mathrm{proto}} < 1$). In Scenarios
1 -- 2, where the giant planet has experienced the lengthiest
migration to $a_\mathrm{g} \lesssim 0.2$~AU, late scattering or
accretion has removed all interior protoplanets (similar to the
results of \citet{fogg3}). In Scenarios 3 -- 5, just one relatively
low mass hot-Earth remains at the end point. The inner disk fares
better in late scenarios (7 -- 9) when migration is limited to
$a_\mathrm{g} \gtrsim 1$~AU. Numerous interior protoplanets remain
in these cases, and since type~I migration forces are absent,
accretion via giant impacts is not suppressed and protoplanetary
growth is more advanced. Since both run sets produce such different
interior partitions, it is of considerable interest to determine if
this effects their final architectures after the subsequent phase of
gas-free accretion. This issue is followed up in
Sect.~\ref{outerplanets}.

Some features of the interior systems illustrated in
Figs~\ref{figure:6}~\&~\ref{figure:7} are worthy of further comment.

\emph{Scenarios 3 -- 5}: Only one interior planet survives in each
of these scenarios where the giant planet strands between
$a_\mathrm{g} = $~0.27 -- 0.55~AU, and only one of these is found at
a first order resonance with the giant at the end point. This is the
$1.90~\mathrm{M}_{\oplus}$ planet in Scenario~4, which is captured
at the 2:1 resonance. In the case of Scenario~3, the interior planet
is found closer to the giant planet, whilst in Scenario~5 the
sweeping 2:1 resonance has not quite reached the surviving
$1.60~\mathrm{M}_{\oplus}$ planet.

\emph{Scenario 6}: The giant planet has stranded at $a_\mathrm{g} =
0.68$~AU, leaving four surviving protoplanets in the interior
partition. The outermost of these has an orbit with $e \approx 0.3$,
close to a 7:4 period ratio with the giant planet, that crosses the
orbit of its nearest neighbor. To an accuracy of $< 1~\%$, period
ratios between the protoplanets, from the outside in, are 4:3, 5:4,
and 3:2 respectively, which are a feature reminiscent of the
resonant convoys of protoplanets commonly observed in simulations
where type~I eccentricity damping is included. In this case,
dynamical friction from surviving planetesimals exerts the damping,
but its relative weakness produces an arrangement that is more
dynamically excited and clearly unstable.

\emph{Scenario 8}: This case stands out from its adjacent Scenarios
7 \& 9, where the giant planet also strands at $a_\mathrm{g} >
1$~AU, because planetary growth in the interior partition appears to
be much more advanced, resulting in three planets in well-spaced
orbits. This is entirely due to chance giant impacts shortly before
the scenario end point. It is shown in Sect.~\ref{outerplanets}
that, if carried through into the gas-free phase, accretion within
the interior partitions of Scenarios 7 \& 9 rapidly catches up, with
excess protoplanets being eliminated by co-accretion or impact onto
the giant planet.

\emph{Scenario~$1_\mathrm{I}$}: The single interior planet resulting
from this run is the best hot-Neptune/super-Earth analogue generated
in this paper. Its mass of $7.57~\mathrm{M}_{\oplus}$ puts it within
the observed range for such objects. However, in contrast with the
results of \citet{fogg4}, the planet is not found at the first order
3:2 or 2:1 resonances, but it is located further away from the giant
planet, at the second order 3:1 resonance (confirmed with a plot of
librating resonant angles). Study of this system's evolution however
reveals that this object was originally shepherded inward at the
2:1, but within the last 0.12~Myr of the run it scatters with and
eventually accretes two other protoplanets interior to it,
contracting the orbit of the merged body, whereupon it is
fortuitously captured at the 3:1 resonance. This phenomenon of
resonance capture through scattering has been recently described by
\citet{raymond4}. The arrangement of hot-Neptune and hot-Jupiter in
Scenario~$1_\mathrm{I}$ is most similar to that observed between the
two inner planets $e$ and $b$ in the 55~Cancri system
\citep{mcarthur}, although the orbits of these two natural objects
are separated by a wider $\sim$~5:1 period ratio.

\emph{Scenario~$2_\mathrm{I}$}: Two interior planets survive in this
case where the giant strands at $a_\mathrm{g} = 0.22$~AU. The
outermost planet of this pair is in the 2:1 resonance with the giant
planet.

\emph{Scenario~$3_\mathrm{I}$}: Two equal mass hot-Earths survive at
the end point of this scenario ($a_\mathrm{g} = 0.33$~AU), but
neither of them are in resonant orbits. A resonant convoy of five
interior planets becomes unstable in the last 40\,000~yr of the
simulation and is broken up by giant impacts.

\emph{Scenario~$9_\mathrm{I}$}: A noteworthy feature of the end
point of this run is the presence of a $0.27~\mathrm{M}_{\oplus}$
Trojan protoplanet, located in a stable co-orbital resonance with
the giant planet. This object was captured into this 1:1 resonance
shortly after the introduction of the giant and escorts the larger
body inward during type~II migration. Although the capture of the
Trojan planet in this case may have been influenced by the abrupt
introduction of the giant planet at the scenario start time, its
presence at the end point is not necessarily unrealistic or
unexpected. \citet{dvorak} \& \citet{erdi} have shown that
co-orbital motions can be stable for long time scales and 1:1
capture and co-migration, in a context where type~I migration forces
are active, has previously been observed in simulations by
\citet{cresswell1,cresswell2}.

\begin{figure}
 \resizebox{\hsize}{!}{\includegraphics{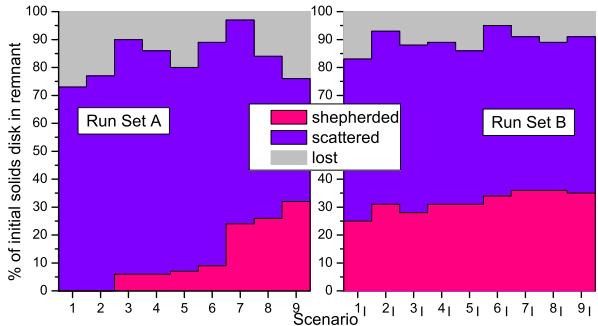}}
 \caption[Fate of the disk mass at the end points of both Run Sets]
 {Fate of the disk mass at the end points of Run Set \textbf{A} \&
 \textbf{B}. Data include the cumulative \% of initial solids
 disk in the shepherded or scattered remnant, or lost from the disk.}
 \label{figure:8}
\end{figure}

\begin{figure}
 \resizebox{\hsize}{!}{\includegraphics{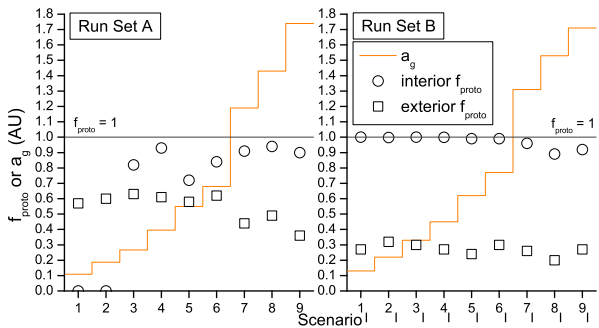}}
 \caption[Protoplanet mass fraction and giant semi-major axis at the
 end points of both Run Sets]
 {Protoplanet mass fraction $f_{\mathrm{proto}}$ interior to giant (circles)
 and exterior to giant (squares). Also shown is the stranded giant planet semi-major
 axis $a_\mathrm{g}$ (stepped line) at the end points of both Run Sets.}
 \label{figure:9}
\end{figure}

Alternate visualizations of the data for the end points of Scenarios
1 -- 9 and Scenarios $1_\mathrm{I} - 9_\mathrm{I}$ are presented in
Figs.~\ref{figure:8}~\&~\ref{figure:9} respectively. These data are
the equivalent of the `fate of the solids disk mass' results of
previous models set out in Tables in the previous papers in this
series and have to be recast into graphical format here since 18
scenarios are under consideration, rather than 5 or 6. The nine
scenarios of each run set are indicated on the x-axis with data from
Run Sets \textbf{A} \& \textbf{B} being shown in the left and right
hand panels respectively. In Fig.~\ref{figure:8}, the cumulative
percentage of the initial solids disk remaining is read off the
y-axis, with the red bars representing the fraction remaining in the
shepherded remnant, the blue bars that in the scattered remnant, and
the grey bars that which is lost -- predominantly via accretion onto
the giant planet. In Fig.~\ref{figure:9}, the y-axis gives both
$f_\mathrm{proto}$ for the interior or exterior remnant (the symbols
being given in the key), and the stranding position of the giant
planet $a_\mathrm{g}$/AU which is shown by the orange line.
Horizontal lines $f_\mathrm{proto} = 0.5$ and $f_\mathrm{proto} = 1$
are drawn for reference.

Fig.~\ref{figure:8} shows that the great majority of the solids disk
($\sim$~80 -- 90~\%) survives giant planet migration, regardless of
whether type~I migration operates or not, and regardless of the
stranding position of the giant planet. The bias toward scattering
behaviour in the absence of type~I migration shows well in the data
for Run Set \textbf{A} as, when $a_\mathrm{g} < 1$~AU, less than
10~\% of the disk mass remains in the shepherded partition. The
biassed scattering of protoplanets is indicated by the value of
exterior $f_\mathrm{proto}
> 0.5$. In contrast, in the presence of type~I migration
(Run Set \textbf{B}), partitioning of the solids disk is largely
insensitive to $a_\mathrm{g}$ with $\sim 30$\% and $\sim 60$\% of
the mass remaining in the interior and exterior partitions
respectively in all scenarios. The bias against the scattering of
protoplanets is indicated by the value of exterior $f_\mathrm{proto}
\approx 0.3$.

One might reasonably speculate, given the details in
Figs.~\ref{figure:6} -- \ref{figure:9}, that the final systems of
terrestrial planets that should emerge from these scenarios might be
either internally or externally weighted (in terms of both planetary
numbers and masses) depending on the strength of type~I migration
forces operating during gas-phase formation. If the model reflects
reality, its results may help contribute to the debate over the
reality of type~I migration once observational techniques have
advanced sufficiently to enable a more complete inventory of
terrestrial planets in exoplanetary systems where type~II giant
planet migration is thought to have occurred.

\subsection{Post-migration terrestrial planet
formation.}\label{outerplanets}

The results of the model, shown in Figs.~\ref{figure:6} \&
\ref{figure:7}, have taken the accretion process only to the point
at which the giant planet strands due to the loss of the nebular
gas. Planet formation in all exterior and most interior partitions
of the original inner system disk is clearly incomplete and awaits a
much lengthier phase of gas-free accumulation. Simulation of at
least part of this gas-free accretion phase is of clear interest to
see if the conclusions drawn from running the model to the end point
of migration might stand up over the long-term. Would a more
complete accretion reduce the observed differences between the two
run sets? What will be the fate of the closely packed resonant
convoys of protoplanets created by type~I migration
(Fig.~\ref{figure:7})? Are habitable zones containing protoplanets
at the end of the migration epoch still likely to contain planets
over the long-term?

\begin{figure*}
\centering
  \includegraphics[width=17cm]{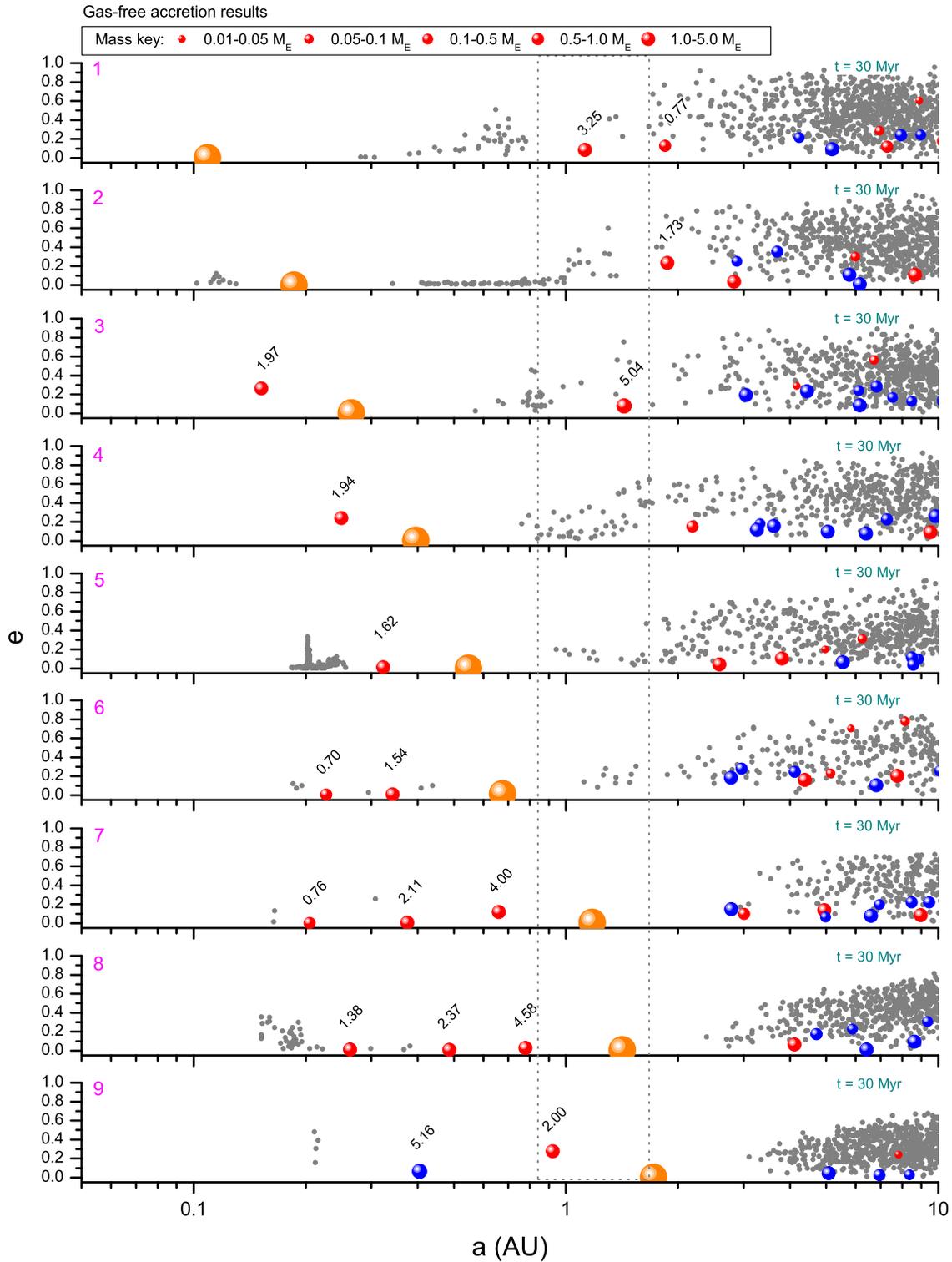}
  \caption{Run Set \textbf{A}. Gas-free accretion results of scenarios
 that \emph{exclude} previous type~I migration. Eccentricity is plotted vs. semi-major
 axis with symbols colour coded as in previous examples and
 sized according to the mass key. Scenario ID is given at the
 top left of each panel. System age is given at the top right of each panel.
 Protoplanets interior to the giant, or within 1 -- 2~AU, are labelled with their
 mass in $\mathrm{M}_\oplus$.  The dotted box shows the habitable zone.}
  \label{figure:10}
\end{figure*}

\begin{figure*}[!]
\centering
  \includegraphics[width=17cm]{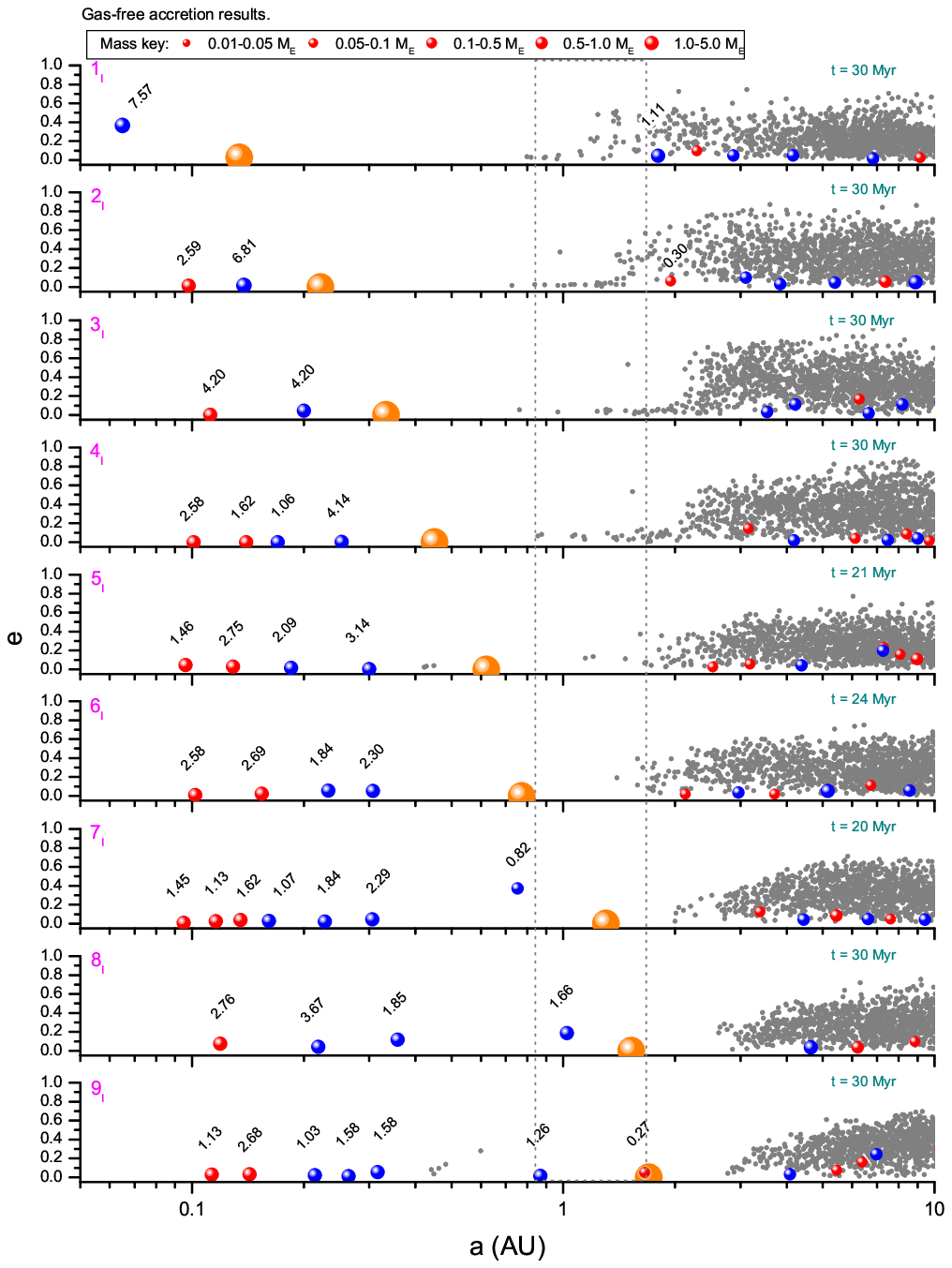}
  \caption{Run Set \textbf{B}. Gas-free accretion results of scenarios that \emph{include} previous type~I
 migration. Scenario ID is given at the top left of each panel. System age is given at the top right of each
 panel. Protoplanets interior to the giant, or within 1 -- 2~AU, are
 labelled with their mass in $\mathrm{M}_\oplus$.  The dotted box shows the habitable zone.}
  \label{figure:11}
\end{figure*}

Thus, after switching off all residual forces exerted by the last
traces of gas, simulation of all the scenarios of both Run Sets has
been extended with the aim of running them to a system age of 30~Myr
($t = 29.5$~Myr). However, this is not practical in every case as
the time-step is constrained by the orbits of the innermost objects.
The method adopted here is to choose a time step equivalent to one
tenth of a circular orbit with a semi-major axis equal to the
periastron of the innermost large object. This strategy typically
dictates a time-step of 2 -- 4 days and serves well in all but two
scenarios. These are Scenario~3 in which a 2 day time-step
inadequately resolves the orbits of a planetesimal swarm interior to
the hot-Earth at 0.18~AU, resulting in an unrealistic loss of these
objects via accretion onto both the star and the hot-Earth; and
Scenario~$1_\mathrm{I}$ where a hot-Neptune at 0.065~AU dictates too
small a time-step -- a problem that is tackled by splitting the
problem into sub runs -- 1) the hot-Neptune and hot-Jupiter with a
0.3 day time-step, 2) the hot-Jupiter and external disk with a 2 day
time-step, and then assembling both as a composite. Even so, eight
months after beginning most of these extended runs it has not been
possible before submission of this paper to complete every
simulation. All scenarios in Run Set \textbf{A} have attained the
goal of 30~Myr, but in Run Set \textbf{B}, where previous type~I
migration has typically resulted in protoplanets closer to the star
at the point of gas loss, the results presented here vary in their
ages from 20 -- 30~Myr.

The results of these extended simulations are illustrated in
Figs.~\ref{figure:10} (Run Set \textbf{A}) \& \ref{figure:11} (Run
Set \textbf{B}) and should be compared with their migration
end-point equivalents in Figs.~\ref{figure:6} \& \ref{figure:7}
respectively. A number of general observations can be made.\\

\noindent 1. \emph{Accretion interior to the giant planet.}
Solitary, or paired, hot-Neptunes or hot-Earths that are present at
the end of the migration epoch (e.g. Scenarios 3 -- 5 and Scenarios
$1_\mathrm{I} - 3_\mathrm{I}$) are found to survive the extended
gas-free accretion phase with little change in their orbital
parameters. The crowded interior disk partitions that result in
later scenarios where the giant comes to rest at greater distances
from the central star undergo rapid evolution to a state of near
completion, with giant impacts thinning down the number of
protoplanets which mop up almost all residual planetesimal debris.

As discussed in Sect.~\ref{modIVres}, giant impact growth in Run Set
\textbf{A} is already underway before the end of the migration phase
as the protoplanets within interior partitions are sufficiently
dynamically excited to exhibit crossing orbits. This process
completes rapidly after gas loss, with protoplanetary mergers
reducing their numbers by $\sim 50\%$, resulting in apparently
stable multiple interior systems of two or three hot-Earths
separated by $> 20$ mutual Hill radii (note Scenarios 6 -- 9,
Fig.~\ref{figure:10}).

\begin{figure}
 \resizebox{\hsize}{!}{\includegraphics{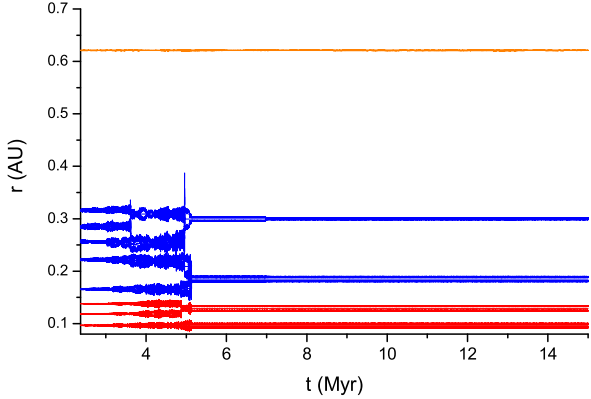}}
 \caption{Destabilization of the resonant convoy in Scenario~$5_\mathrm{I}$. Temporal
 evolution of the periastron, semi-major axis, and apastron for each object
 is shown, with the giant planet drawn in orange at the top of the graph.
 Protoplanets undergo a sequence of closest neighbour mergers, reducing their
 number from eight down to a stable quartet of hot-Earths.}
 \label{figure:12}
\end{figure}

\begin{figure}
 \resizebox{\hsize}{!}{\includegraphics{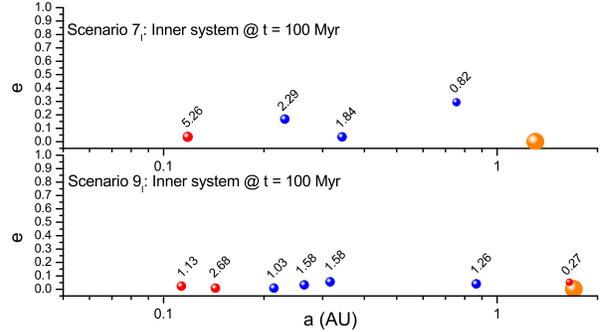}}
 \caption{Inner systems of Scenarios~$7_\mathrm{I}$~\&~$9_\mathrm{I}$ run to
 $t = 100$~Myr.}
 \label{figure:13}
\end{figure}

In Run Set \textbf{B}, type~I migration forces have previously
suppressed giant impact growth, causing protoplanets to stack into
crowded resonant convoys (e.g. Scenarios~$5_\mathrm{I} -
9_\mathrm{I}$, Fig.~\ref{figure:7}). After the gas is gone in these
cases, all eccentricity damping ceases and the convoys start to
destabilize within a few Myr, typically by protoplanets merging with
their nearest neighbours, ultimately reducing their numbers by $\sim
50\%$. This occurs fastest in convoys that are the most compressed
by the position of the giant planet (i.e. restricted within the
narrowest annulus) and the process is clearly illustrated in the
case of Scenario~$5_\mathrm{I}$ in Fig.~\ref{figure:12}. Generally
however, systems where strong type~I migration has operated in the
gas phase still retain more interior planets, although is it not
certain that all giant impact growth is completed within the
interior partitions of the later scenarios illustrated in
Fig.~\ref{figure:11}. Scenarios~$7_\mathrm{I}~\&~9_\mathrm{I}$ have
interior protoplanets still separated by as little as $\sim 10$
mutual Hill radii and some traces of the original resonant convoy
structure remains, such as the 4:3, 5:4, 4:3 commensurability
between the inner four planets of Scenario~$7_\mathrm{I}$. To test
the stability of the still crowded inner systems shown in
Fig.~\ref{figure:11}, the orbits of the giant and the remaining
interior planets in Scenarios~$7_\mathrm{I}$~\&~$9_\mathrm{I}$ were
integrated to $t = 100$~Myr and the results are displayed in
Fig~\ref{figure:13}. It can be seen that whilst the inner system of
seven planets in Scenario~$9_\mathrm{I}$ is stable over this
extended time frame (including the orbit of the Trojan planet), that
of Scenario~$7_\mathrm{I}$ undergoes further orbital evolution and
accretion resulting in a system of four survivors. Whatever the
final outcome in these cases however, hot and warm-Earths, including
multiple systems, are predicted as an outcome of type~II migration
induced disk compaction in all but the earliest scenarios
of Run Set \textbf{A}.\\

\noindent 2. \emph{Accretion exterior to the giant planet.}
Planetary growth in exterior partitions at $\gtrsim 2$~AU is ongoing
and not yet complete. This is to be expected as dynamical times are
longer, scattered material is spread over a larger volume, and does
not have an opportunity to interact with disk solids originating
beyond 5~AU which have not been modelled. However, it can be seen in
Figs.~\ref{figure:10} \& \ref{figure:11} that dynamical friction
exerted by the scattered planetesimal population has resulted in a
degree of circularization of the orbits of scattered protoplanets,
especially in the case of Run Set \textbf{B}, where dynamical
friction is particularly strong due to the low exterior value of
$f_\mathrm{proto}$ (see Fig.~\ref{figure:9} and associated
discussion).\\

\noindent 3. \emph{Planetary occupants of the habitable zone.} Three
planets in each run set are found in their system's habitable zone
($\sim 0.84 - 1.67$~AU) after the extended gas-free runs. In Run Set
\textbf{A}, these are 3.25, 5.04 \& 2.00~$\mathrm{M}_\oplus$ planets
in Scenarios 1, 3 \& 9 respectively (Fig.~\ref{figure:10}); and in
Run Set \textbf{B} a 1.66~$\mathrm{M}_\oplus$ planet in
Scenario~$8_\mathrm{I}$ and 1.26 \& 0.27~$\mathrm{M}_\oplus$ planets
in Scenario~$9_\mathrm{I}$ (Fig.~\ref{figure:11}). This latter
object is the Trojan planet captured into a 1:1 resonance with the
giant early on in the migration phase (discussed in
Sect.~\ref{modIVres}) and dragged inward to $a = 1.66$~AU, just
inside the outer edge of the habitable zone. This exotic world
remains in a continuously stable orbit up to a system age of 30~Myr,
and has the potential for a prolonged existence
\citep[][\&~Fig.~\ref{figure:13}]{dvorak,erdi}. The finding
articulated in the previous Section that the giant planet must make
either a limited excursion into the habitable zone, or a complete
traversal down to $\lesssim 0.5~\times$~the radial distance of the
inner edge of the HZ, for a habitable planet to be possible is
reinforced by the results of the extended simulations. Potentially
habitable planets are found in systems where the giant planet lies
outside the region of $\sim 0.3 - 1.5$~AU, except in the case where
a Trojan planet accompanies a giant stranded within the HZ.\\

\begin{figure}
 \resizebox{\hsize}{!}{\includegraphics{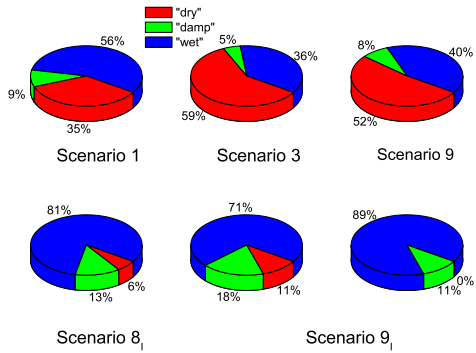}}
 \caption
 {Material composition of habitable planet candidates from our
 simulations. Red
 indicates material of provenance $a < 2$~AU; green indicates material of
 provenance $2~\mathrm{AU} < a < 2.7~\mathrm{AU}$; and blue indicates
 trans-snowline material originating from $a > 2.7$~AU. The top row of
 charts are the results of Run Set \textbf{A} and the bottom charts the results of
 Run Set \textbf{B}.}
 \label{figure:14}
\end{figure}

\noindent 4. \emph{Volatile endowments.} In agreement with previous
models \citep{fogg3,fogg4} and the work of \citet{raymond2} and
\citet{mandell}, the habitable planets generated in the present
model are predicted to be richly endowed with volatiles driven
inward from beyond the nebular snowline by the migrating giant
planet. The composition of the six habitable planet candidates,
using the same crude three phases described in \citet{fogg3}, are
shown as pie charts in Fig.~\ref{figure:14} where the red sectors
refer to dry material originating interior to 2~AU, the green
sectors to chondritic material originating between 2 and 2.7~AU, and
the blue sectors indicate the fractions originating from beyond the
snowline at $> 2.7$~AU. Even with no previous type~I migration
operating (Run Set \textbf{A}), the habitable planet candidates
mentioned above have incorporated roughly a third to a half of their
material from beyond the snowline. Where there has been pre-existing
type~I migration (Run Set \textbf{B}), more like three quarters of
the planets' mass originates from beyond the snowline. Whether the
habitable planets in question lie interior or exterior to the giant
seems to make little difference to the composition. All such worlds
are predicted to be richly endowed with water and gases.

\section{Discussion.}

The key finding of the model presented in this paper, and its
predecessors \citep{fogg1,fogg2,fogg3,fogg4}, that giant planets do
not remove the portion of a protoplanetary disk they migrate
through, or prevent subsequent terrestrial planet formation, is
robust in the sense that every model variant, and independent work
\citep{raymond2,mandell}, has reproduced it. However, these models
have only explored a modest region of parameter space relevant to
the problem, and have inevitably adopted assumptions that simplify
or omit potentially important physical processes. We have discussed
a number of issues concerning parameter variations and neglected
physical processes in our previous papers and restrict the
discussion here to three caveats not previously considered.\\

\noindent \emph{i) Solids disk inner boundary}. The inner system
solids disk adopted here begins with its inner boundary set at
0.4~AU. Its contents are then free to evolve inward during the
maturation runs preceding the introduction of the giant at the start
of a migration scenario. By this time, in low dissipation models
(Run Set \textbf{A}), dynamical spreading typically results in a
modest inward movement of the disk edge to $\sim 0.3$~AU (see the
upper panel of Fig.~\ref{figure:4}) so that when the giant planet
subsequently migrates down to 0.1~AU it passes through the entire
inner disk annulus. This is not necessarily the case in high
dissipation models (Run Set \textbf{B}) where greater inward
movement of the solids disk contents occurs during the maturation
phase (see the lower panel of Fig.~\ref{figure:4}). Thus, the
question is raised as to whether the lack of interior hot-Earths
resulting in Scenarios~1 \& 2 is a realistic finding or an artifact
of the initial position of the solids disk inner boundary. What if
this boundary had been set at 0.1~AU instead? This would have
provided an additional $\sim 3~\mathrm{M_\oplus}$ of material from
which potential interior planets could have been assembled. It
cannot be ruled out therefore that some hot-Earths could grow and
survive even in early scenarios of Run Set \textbf{A}, especially if
the giant planet stops short of their position. However, one would
still expect them to be of lower mass and more prone to orbital
destabilization than in high dissipation models where eccentricity
damping is in play and where there has been much more extensive
prior inward migration of solids.\\

\noindent \emph{ii) Random component of initial conditions}. The
solids disk with which we begin our simulations at $t = 0$ conforms
as a whole to a 3$\times$MMSN model, but is unique in the sense that
its individual particles start with randomized orbital elements.
This raises the question of how the conclusions of this paper might
depend on the seed number used to randomize these initial
conditions. We have not tested the influence of alternate seed
numbers on the particular set of simulations presented here.
However, comparison with our previous work, where different integers
have been used to seed the initial setup, shows that in scenarios
where disk maturity and dissipative forces are roughly equivalent,
the overall architectures of the post-migration systems are similar
even though details of individual planets are different. This is
because the general physical processes shown to characterize the
migration of a giant planet through an inner system disk, which lead
to the division of the original disk contents into internal and
external remnants, are largely independent of the random component
in the initial conditions and are influenced primarily by the level
and variety of dissipative forces in play -- i.e. by the maturity of
the solid body swarm, the density of the gas, and whether type~I
migration is active. An example of such a comparison can be seen by
referring to Fig.~12 from \citet{fogg3}, where the results of a late
scenario, matured for 1.5~Myr before the onset of giant planet
migration, in the absence of type~I migration, are plotted. The
lower panel of this figure shows the evolution of this model 2~Myr
after the giant planet has stranded at 0.1~AU and its overall system
architecture is similar to its closest equivalent in the present
work: Scenario~1, shown in the top panel of Fig.~\ref{figure:6}. No
interior hot-Earth is present and protoplanets amounting to several
$\mathrm{M}_{\oplus}$ have been externally scattered to between 1 -
3~AU in both cases. The post migration evolution of this hot-Jupiter
system from \citet{fogg3} has been run further to 30~Myr and the
results illustrated in \citet{fogg5}. The resemblance between that
run and the matured Scenario~1 in this paper (Fig.~\ref{figure:10},
top panel) is even closer with a $\sim 3.3~\mathrm{M}_{\oplus}$
planet being present in the habitable zone of both systems. We
conclude that although the specific properties of the planetary
systems which arise from our simulations depend on the exact details
of the initial setup, the general trends that we observe in our
outcomes are robust. \\

\noindent \emph{iii) Ocean planets}. One effect of giant planet
migration is to drive large quantities of icy material into the
inner system, with the outcome that terrestrial planets that form in
the aftermath are predicted to be richly endowed with volatiles and
enveloped by global oceans. There is nothing about 100\% ocean cover
that necessarily rules out the presence of life. Indeed, life may
have started in the Earth's oceans, and exploration beneath
ocean-covered Europa's ice shell has long been regarded as a top
priority by astrobiologists \citep{reynolds1,reynolds2}. Ocean
planets in the $\sim \mathrm{M}_\oplus$ range however have a problem
not shared with Europa: if the depth of their oceans is $\gtrsim
100$~km, their floors are composed of high pressure phases of ice,
rather than rock, with the ice-silicate interface (and much
potentially crucial chemistry) sealed off beneath an icy mantle 100s
or 1000s of km thick, depending on the planet's total mass and bulk
water content \citep{leger}. How this would effect the solute
content of the oceans is unknown but one might speculate that some
trace elements essential for terrestrial life might be
lacking\footnote{The elemental requirements of
\emph{non-terrestrial} life are another matter and unknown.}.
Meteoritic infall after planetary differentiation is complete would
re-supply some rock-forming elements for dissolution into the ocean,
but solid material that deposits on the ocean floor would be denser
than the ice beneath and prone to removal by sinking into the
mantle. A related problem deepens this uncertainty. Planets
over-endowed with water might similarly possess massive atmospheres
with unexpected compositions and properties. If ocean planets are
more like thawed out versions of Titan, rather than water-rich
Earths, then habitable zone calculations that rely on models of
modest $\mathrm{CO_2} / \mathrm{H_2O}$ atmospheres, coupled to a
carbonate-silicate cycle and continental weathering
\citep[e.g.][]{kasting,selsis,vonBloh}, may not be relevant.

If the water content of trans-snowline planetary building blocks is
assumed to be $\sim 75$\% ($\sim 1 - f_\mathrm{ice}^{-1}$ where
$f_\mathrm{ice} = 4.2$ is the factor in the MMSN model by which
nebular condensate masses are enhanced due to ice condensation),
then the models presented here predict HZ planets composed of $\sim
20$\% -- 60\% water. This is a vast inventory compared with the
Earth's $\sim 0.1$\% water. It is likely though that these figures
are overestimated. If the snowline is further from the star than the
2.7~AU assumed in the MMSN~Model, or if the giant planet forms
closer to it, the material shepherded inward by type~II migration
will have a lower volatile content. In the models of
\citet{raymond2} and \citet{mandell}, where the snowline is placed
at 5~AU, HZ planets are produced with a water content of $\sim 10$\%
-- a reduction from the above estimate, but still a factor of 20
greater than the typical outcome of $\sim 0.5$\% water content when
models are run without a migrating giant planet
\citep{raymond1,raymond3}. However, all these models almost
certainly overestimate the quantity of water retained by growing
planets since loss of volatiles during accretion is not accounted
for. Extensive depletion of both atmosphere and ocean could result
from giant impacts late in formation \citep{genda,asphaug,canup}
which could strip nascent ocean planets down to a more Earth-like
remnant. All of the HZ planets generated by the models presented
here undergo \emph{at least} one giant impact at some point in their
evolution, with a minority suffering a high velocity collision with
a comparable sized object during the later stage of accretion within
the scattered disk. Thus, whilst ocean planets in habitable zones
are a robust prediction of the model, it is premature to say that
they would invariably occur in nature and would inevitably be
uninhabitable.

\section{Conclusions.}

All previous published models of terrestrial planet formation in the
presence of type~II giant planet migration have neglected the issue
of what causes the giant planet to come to rest in its final orbit
and have simply switched off migration forces at a pre-defined
distance from the central star
\citep{fogg1,fogg3,fogg4,raymond2,mandell}. This latest model is the
first simulation of this type to include a self-consistent scenario
for the stranding of the giant planet, which comes to rest naturally
at the point when the gaseous fraction of the protoplanetary disk is
lost via a combination of accretion onto the central star and
photoevaporation.

The results of the earliest scenarios -- those which simulate
hot-Jupiter emplacement at $a_\mathrm{g}\approx 0.1$~AU -- are very
similar to the late scenarios of \citet{fogg3,fogg4}, where the
evolution of the nebula is advanced and dissipative forces that stem
from the presence of gas are relatively weak. Most of the inner
system disk solids survive the traverse of the giant planet and most
of this surviving matter is scattered into higher orbits where
planet formation can resume. Water-rich habitable planets are
possible within a habitable zone that is far removed from the final
position of the giant planet. Where dissipative processes are
enhanced by type~I migration, one or two hot-Earths or hot-Neptunes
are found to persist in orbits interior to the giant planet.

In `warm-Jupiter' type scenarios, where $a_\mathrm{g} \gg 0.1$~AU,
the bulk of the disk solids survive as before, but more complex
systems of hot terrestrial planets interior to the giant are
predicted, in larger numbers with increasing $a_\mathrm{g}$. If
strong type~I migration forces are a genuine influence on planet
formation, relatively crowded systems of interior planets are
predicted, although it has not been practical to check this forecast
with Gyr-long integrations. Habitable planets in low eccentricity
warm-Jupiter systems appear possible if the giant planet makes a
limited incursion into the outer regions of the habitable zone, or
traverses its entire width and keeps going until it ceases migrating
at a radial distance of less than half that of the HZ's inner edge:
i.e. the giant planet's final orbit lies outside the region of $\sim
0.3 - 1.5$~AU for a 1~$\mathrm{M_\odot}$ star. If giant planets
strand within this region, the presence of habitable planets remains
a possibility if they are located in a stable 1:1 resonance with the
giant. These findings should hold for stars with different
luminosities with an appropriate scaling of radial distance.

\begin{figure}
 \resizebox{\hsize}{!}{\includegraphics{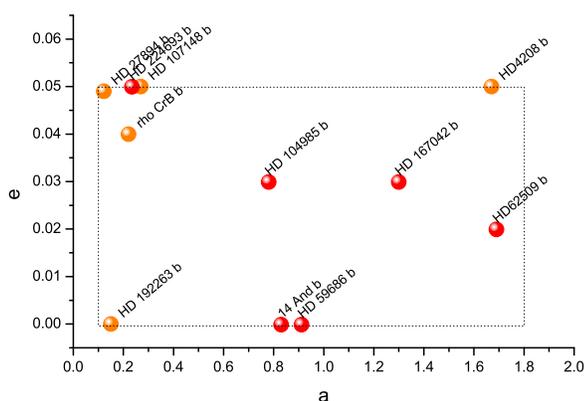}}
 \caption{Data of low eccentricity exoplanets
 using data from the Extrasolar Planets
 Encyclopedia: http://exoplanet.eu. Plotted objects correspond to
 the type of warm-Jupiters within the remit of the model presented here: solitary giant planets
 orbiting single stars with $0.1~\mathrm{AU} < a < 1.8$~AU, $e < 0.05$.
 Orange symbols represent exoplanets within main sequence star systems; red symbols
 represent exoplanets orbiting post-main sequence stars.}
 \label{figure:15}
\end{figure}

However, warm-Jupiter type exoplanets with near-circular orbits,
i.e. $e \lesssim 0.05$, are relatively uncommon and $e$ is typically
observed to be considerably higher. Many of these exoplanets may
have had an origin involving mutual giant planet scattering, perhaps
combined with migration \citep[e.g.][]{ford,papaloizou0b,moorhead1},
as opposed to the strongly damped pure type~II migration mechanism
adopted here. Nevertheless, low-$e$ and solitary warm-Jupiters are
known, which presumably could have originated in a similar manner to
that simulated here. Fig.~\ref{figure:15} uses data from the
Extrasolar Planets Encyclopedia to plot exoplanetary data between $a
= 0.1 - 1.8$~AU vs. $e = 0 - 0.05$ which might be regarded as lying
within the remit of the model. Solitary giant planets around single
stars are plotted and labelled with their names. Four of these
exoplanets, $\rho~\mathrm{CrB~b}$, HD~27894~b, HD~192263~b and
HD~4208~b occur in systems where there is dynamical room for
terrestrial planets in the habitable zone according to the stability
calculations of \citet{jones1,jones2}. Systems containing the
exoplanets HD~224693~b \citep{jones2}, HD 104985~b and HD 59686~b
\citep{jones1} could also have hosted planets in their habitable
zones in the past before their primaries left the main sequence.
According to the results of the model presented in this paper, it
appears feasible that habitable planets could have originated in
these systems as well.



\bibliographystyle{aa}

\listofobjects

\end{document}